\documentclass[a4paper,10pt]{article}
\usepackage[dvips]{graphicx}
\usepackage{amssymb,amsmath}
\numberwithin{equation}{section}

\begin{document}

\title{Pure kinetic k-essence as the cosmic speed-up}
\author{P.Yu.Tsyba, I.I. Kulnazarov,  K.K. Yerzhanov, R. Myrzakulov\footnote{The corresponding author. Email: rmyrzakulov@csufresno.edu; rmyrzakulov@gmail.com} \\ \textit{Eurasian International Center for Theoretical Physics and  Department of General } \\ \textit{ $\&$  Theoretical Physics, Eurasian National University, Astana 010008, Kazakhstan}}

\date{}

\maketitle

\begin{abstract}
In this paper, we consider three types of k-essence. These k-essence models were presented in the parametric forms. The  exact analytical solutions of the corresponding equations of motion are found. It is shown that these k-essence models for the presented solutions  can give rise to cosmic acceleration. 
\end{abstract}
\vspace{2cm} \sloppy


\section{Introduction}

Dark energy has become one of the most important issues of the modern cosmology ever since the observations of type Ia supernovae (SNe Ia) first indicated that the universe is undergoing an accelerated expansion at the present stage  \cite{Ries}. However, hitherto, we still know little about dark energy.  Dark energy is almost equally distributes in the Universe, and its pressure is negative. The simplest and
most theoretically appealing candidate of dark energy is  vacuum energy (or a cosmological constant $\Lambda$) with a
constant equation of state (EoS) parameter $w = -1$. This scenario is in general agreement with the current astronomical
observations, but has difficulties to reconcile the small observational value of dark energy density with estimates
from quantum field theories; this is the cosmological constant problem \cite{1}. Recently it was shown that $\Lambda$CDM model
may also suffer an age problem \cite{2}. It is thus natural to pursue alternative possibilities to explain the mystery of
dark energy. Over the past decade numerous dark energy models have been proposed, such as quintessence, phantom,
k-essence, tachyon, (generalized) Chaplygin gas, DGP, etc. K-essence, a simple approach toward constructing a model
for an accelerated expansion of the Universe, is to work with the idea that the unknown dark energy component is
due exclusively to a minimally coupled scalar field $\phi$ with non-canonical kinetic energy which results in the negative
pressure \cite{3}.
The k-essence scenario has received much attention, it was originally proposed as a model for inflation \cite{4}, and then as
a model for dark energy \cite{3}. Its action has the form
\begin {equation}
S=\int dx^{4}\sqrt{-g}[\frac{1}{2k^2}R+K(X, \phi)+L_m].
\end{equation}
This general field Lagrangian includes as particular cases the common canonical scalar field model, DBI-inflaton \cite{44} and k-inflation/k-essence \cite{4}.  In several cases, k-essence cannot be observationally distinguished from quintessence \cite{5}.
A method to obtain a phantom version of FRW k-essence cosmologies was devised in [6]. The stability of k-essence
was studied in \cite{7}. Dynamics of k-essence were discussed in \cite{8}. Conditions for stable tracker solutions for k-essence
in a general cosmological background were derived in \cite{9}. Slow-roll conditions for thawing k-essence were obtained
in \cite{10}. A connection between the holographic dark energy density and the kinetic k-essence energy density was
discussed in \cite{11}. An holographic k-essence model of dark energy was proposed in \cite{12}. The geometrical diagnostic
for purely kinetic k-essence dark energy was discussed in \cite{13}. The equivalence of a barotropic perfect fluid with a
k-essence scalar field was considered in \cite{14}, \cite{15}. Dark matter to dark energy transition in k-essence cosmologies was
examined in \cite{16}. Models of dark energy with purely kinetic multiple k-essence sources that allow for the crossing of
the phantom divide line were investigated in \cite{17}- \cite{18}. The thermodynamic properties of  k-essence was discussed in
\cite{19}. Models of k-essence unified dark matter were discussed in \cite{20}-\cite{24}.  Power-law expansion in k-essence cosmology
was investigated in \cite{25}. Theoretical and observational constraints on k-essence dark energy models were discussed
in \cite{26}-\cite{28}. In Ref. \cite{29}, a model independent method of reconstructing the Lagrangian for the k-essence field by
using three parametrizations for the Hubble parameter $H(z)$ was studied in detail. With assumptions on the EoS of
k-essence as functions of the scale factor, Ref. \cite{30} discussed the forms of the Lagrangians. By assuming the EoS of
k-essence as functions of the kinetic energy, Ref. \cite{31} consider restrictions on purely kinetic k-essence. Some properties of k-essence have been studied in \cite{Yang1}-\cite{BBM}.

Since from it was proposed, k-essence was been studied intensively. It is still worth investigating  in a systematic way the possible cosmological behavior of the k-essence cosmology. Thus, in the present work we reconstruct some of k-essence models. We are interested in investigating the possible late-time solutions. In these solutions we calculate various observable quantities, such as the density of the dark energy and the equation of state parameters. As we see,  k-essence cosmology can be consistent with observations. 

This paper is organized as follows. In the following section, we briefly review the model of k-essence. In Sec.III, we study the  $\tanh$-models of k-essence and in Sec. IV, we investigate   their cosmological behaviors.  Finally,  the last section is devoted to  conclusions. Note that in this paper, we have used the metric signature $(-,+,+,+)$ and natural units with $8\pi G=c=\hbar=1$.
 
\section{Briefly review on purely kinetic k-essence}

Let us start from  the following action of pure kinetic k-essence [4, 5, 33]
\begin {equation}  
S=\int dx^{4}\sqrt{-g}[\frac{1}{2k^2}R+K(X)+L_m],
\end{equation}
where  $L_m$ is  the  matter Lagrangian density. Here $K(X)$ is the k-essence Lagrangian, which depends  on the dimensionless quantity $X$ defined by
\begin {equation}
X=-0.5g^{\mu\nu}\phi_{\mu}\phi_{\nu}.
\end{equation}
A possible motivation for actions of this form comes from considering low-energy effective string theory in the presence of a high-order derivative terms \cite{33}-\cite{35}. Such theories have been exploited as models for inflation and dark matter/energy, for example a purely kinetic k-essence  or ghost condensate. The simplest nontrivial example of purely kinetic k-essence is a ghost condensate Lagrangian 
\begin {equation}
K_{gh}=\alpha(1-X)^2,
\end{equation}
which has been also studied in the context of dark matter/energy unification. Another example originates from the Dirac-Born-Infeld description of a $d$-brane in string theory with the scalar Born-Infeld  type Lagrangian 
\begin {equation}
K_{DBI}=\alpha\sqrt{1-X}.
\end{equation}

The energy momentum tensor obtained from (2.1) takes the perfect fluid form,
\begin {equation}
T_{\mu\nu}=-\frac{2}{\sqrt{-g}}\frac{\delta(\sqrt{-g}K)}{\delta g^{\mu\nu}}= 2K_{X}\phi_{\mu}\phi_{\nu}+Kg_{\mu\nu}=(\rho_k+p_k)u_{\mu}u_{\nu}+pg_{\mu\nu}.
\end{equation}
Here
\begin {equation}
u_{\mu}=\eta\frac{\phi_{\mu}}{\sqrt{X}},
\end{equation}
where $\eta$ is $+1$ or $-1$ according to whether $\phi_{0}$ is positive or negative, i.e., the sign of of $u_{\mu}$ is chosen so $u_{0}$ is positive. The associated hydrodynamic quantities are
\begin {equation}
p_k=K, \quad \rho_k=2XK_{X}-K
\end{equation}
so that the equation of state parameter, $w$, is
\begin {equation}
w_{k}=\frac{p_k}{\rho_k}=\frac{K}{2XK_{X}-K}.
\end{equation}
Here $p_k, \rho_k$ are the pressure of the scalar field $\phi$ and its energy density, respectively.
We also introduce the sound speed of k-essence, which is the relevant quantity for the growth of density perturbations,
\begin {equation}
c_{s}^{2}=\frac{p_{kX}}{\rho_{kX}}=\frac{K_{X}}{2XK_{XX}+K_{X}}.
\end{equation}
Let us rewrite Eq.(2.9) as
\begin {equation}
2XK_{XX}+(1-c_{s}^{-2})K_{X}=0.
\end{equation}
The definition of the sound speed comes from the equation describing the evoluion of linear adiabatic perturbations in a k-essence dominated Universe \cite{33} (the non-adiabatic perturbation was discussed in \cite{36}-\cite{38}, here we only consider the case of adiabatic perturbations). Perturbations can become unstable if the sound speed is imaginary, $c^2_{s}<0$, so we insist on $c^2_{s}\geq0$. Another potentially interesting requirement to consider is $c^2_{s}\leq 1$, which says that the sound speed should not exceed the speed of light, which suggests violation of causality. Note, however, this is still an open problem (see e.g. \cite{39}-\cite{44}).
We now consider the spatially flat Friedmann-Lemaitre-Robertson-Walker (FLRW) space-time with a time-dependent scale factor $a(t)$ and a metric 
\begin {equation}
ds^{2}=-dt^{2}+a(t)^{2}\sum^{3}_{i=1}(dx^{i})^{2},
\end{equation}
where $t$ is  cosmic time. For this metric the equations of motion and the continuty equation take the form
\begin{align}
\frac{3}{k^2}H^2=2XK_{X}-K+\rho_m, 
\end{align}
\begin{align}
	-\frac{1}{k^2}(2\dot{H}+3H^2)=K+p_m, 
\end{align}
\begin{align}
(K_X+22XK_{XX})	\dot{X}+6HXK_X=0,
\end{align}
\begin{align}
	\dot{\rho_m}+3H(\rho_m+p_m)=0.
\end{align}
As known, from (2.14) or from its equivalent form $\dot{\rho}_k+3H(\rho_k+p_k)=0$ follows that 
\begin{align}
	XK_X^2=\kappa a^{-6}, 
\end{align}
where $\kappa=const\leq 0$ since $X=0.5\dot{\phi}^2$. Also we note that as $K(X)=0$ Eqs. (2.12)-(2.13) transform to the usual Friedmann equations of GR
\begin{equation}
\frac{3}{k^2}H^2=\rho_m, \quad \frac{1}{k^2}(2\dot{H}+3H^2)=-p_m.
\end{equation}
So we can rewrite the equations (2.12)-(2.13) as
\begin{equation}
\frac{3}{k^2}H^2=\rho_m+\rho_k=\rho_{tot}, \quad \frac{1}{k^2}(2\dot{H}+3H^2)=-(p_m+p_k)=-p_{tot},
\end{equation}
where $\rho_k,  p_k$ are the k-essence  contributions to the energy density and pressure. We also present the parameter of state 
\begin{align}
	w_{tot}=\frac{p_{tot}}{\rho_{tot}}=\frac{p_k+p_m}{\rho_k+\rho_m}=-1-\frac{2}{3}(\ln{H})_{N},\quad w_{k}=\frac{p_k}{\rho_k}, \quad w_m=\frac{p_m}{\rho_m}.
\end{align}

\section{New types of k-essence}

Let us introduce the following form of the k-essence
\begin{align}
	K=\sum_{j=0}^{M}\nu_jy^j,
\end{align}
where $y=\tanh[t]$ and in general $\nu_j=\nu_j(\phi)=\nu_j(t)$. But in this work we restrict ourselves to the case when  $\nu_j=const.$
The models of the form (3.1) are may be a new types of  of k-essence.  To solve  the gravitional equations we use the reconstruction method (see e.g. \cite{N1}-\cite{MR2}). 
For example,  we can look for $H$ as
\begin{align}
	H=\sum_{j}\mu_jy^{j},
\end{align}
where  in general $\mu_j=\mu_j(t)$. But in this paper we assume that $\mu_j=const$. Below we would like to construct some exact solutions of the equations  (2.12)-(2.13). For simplicity we assume that $\rho_m=p_m=0=k^2-1$. Here also we mention that as we assumed in the end of the introduction section, we work with the natural units where $8\pi G=c=\hbar=1$ so that all variables are dimensionless. In particular, the variables (constants) $\nu_j$ and $\mu_j$ are dimensionless.   Eqs.(2.12)-(2.13) take the form
\begin{align}
3H^2=2XK_{X}-K, 
\end{align}
\begin{align}
	-(2\dot{H}+3H^2)=K. 
\end{align}
Note that 
\begin{align}
\dot{H}=-XK_X 
\end{align}
and
\begin{align}
X=\kappa^{-1}\dot{H}^2a^6. 
\end{align}

\subsection{Example 1: $M=2$}

First we consider the case: $ M=2$. Then
\begin{align}
K=\nu_0+\nu_1 y+\nu_2 y^2.
\end{align}
Let $H$ has the form
\begin{align}
	H=\mu_0+\mu_1y,
\end{align}
so that 
\begin{align}
a=a_0e^{\mu_0t}\cosh^{\mu_1}[t].
\end{align}
From (3.4) we obtain the following explicit form of the k-essence Lagrangian
\begin{align}
	K=-(2\mu_1+3\mu_0^2)-6\mu_0\mu_1y+3\mu_1(2-\mu_1)y^2.
\end{align}
At the same time, Eq.(3.3) we gives
\begin{align}
2XK_X=3H^2+K=-2\dot{H}=-2\mu_1(1-y^2).
\end{align}
For $X$ we get the following expression 
\begin{align}
X=\kappa^{-1}a_0^6\mu_1^2e^{6\mu_0t}\cosh^{4+6\mu_1}[t]=\kappa^{-1}a_0^6\mu_1^2e^{2st}\cosh^{2l}[t].
\end{align}
The corresponding scalar function $\phi$ is
\begin{align}
\phi=\partial^{-1}_t\sqrt{-2X}=\phi_0-\sqrt{-2\kappa^{-1}}a_0^3\mu_1\frac{2^{-l}e^{st}(e^{-t}+e^{t})^l(1+e^{2t})^{-l}Z}{h},
\end{align}
where $Z=Hypergeometric2F1[\frac{h}{2},-l,\frac{1}{2}(2+h),-e^{2t}], \quad l=2+3\mu_1, \quad s=3\mu_0, \quad h=sLog[e]-l,\quad \phi_0=const$. Finally we get the following expressions for the density of energy and pressure:
\begin{align}
\rho_k=3(\mu_0+\mu_1y)^2, \quad
p_k=-(2\mu_1+3\mu_0^2)-6\mu_0\mu_1y+3\mu_1(2-\mu_1)y^2.
\end{align}

\subsection{Example 2: $M=6$}

Here we assume that $M=6$. Then
\begin{align}
	K=\nu_0+\nu_1y+\nu_2y^2+\nu_3y^3+\nu_4y^4+\nu_5y^5+\nu_6y^6.
\end{align}
The Hubble parameter we give in the form
\begin{align}
	H=\mu_0+\mu_1y+\mu_3y^3.
\end{align}
Hence for the scale factor we obtain
\begin{align}
a=a_0\cosh^{(\mu_1+\mu_3)}[t]e^{\{\mu_0t+0.5\mu_3\cosh^{-2}[t]\}}.
\end{align}
Using 
\begin{align}
	\dot{H}=\mu_1+(3\mu_3-\mu_1)y^2-3\mu_3y^4,
\end{align}
\begin{align}
	H^2=\mu_0^2+2\mu_0\mu_1y+\mu^2_1y^2+2\mu_0\mu_3y^3+2\mu_1\mu_3y^4+\mu_3^2y^6
\end{align} 
from (3.4) we get
\begin{align}
	\nu_0: &=&-(2\mu_1+3\mu_0^2),\\
	\nu_1: &=&-6\mu_0\mu_1,\\
	\nu_2: &=&-2(3\mu_3-\mu_1)-3\mu_1^2,\\
	\nu_3: &=&-6\mu_0\mu_3,\\
	\nu_4: &=&6\mu_3(1-\mu_1),\\
	\nu_5: &=&0,\\
	\nu_6: &=&-3\mu_3^2.
	\end{align}
From (3.6) we obtain
\begin{align}
X=\frac{a_0^6}{\kappa}[\mu_1+(3\mu_3-\mu_1)y^2-3\mu_3y^4]^2e^{[6(\mu_0t+0.5\mu_3\cosh^{-2}[t])]}\cosh^{6(\mu_1+\mu_3)}[t].
\end{align}

The density of energy and pressure are given by
\begin{align}
\rho_k=3(\mu_0+\mu_1y+\mu_3y^3+\mu_5y^5)^2,
\end{align}
\begin{align}
p_k=\nu_0+\nu_1y+\nu_2y^2+\nu_3y^3+\nu_4y^4+\nu_5y^5+\nu_6y^6,
\end{align}
respectively.
\subsection{Example 3: $M=10$}

Finally we consider the example when $M=10$ so that the corresponding k-essence has the form
\begin{align}
	K=\nu_0+\nu_1y+\nu_2y^2+\nu_3y^3+\nu_4y^4+\nu_5y^5+\nu_6y^6+\nu_7y^7+\nu_8y^8+\nu_9y^9	+\nu_{10}y^{10}.
\end{align}
We assume that $H$ given in the form 
\begin{align}
	H=\mu_0+\mu_1y+\mu_3y^3+\mu_5y^5,
\end{align}
so that for the scale factor we have the expression
\begin{align}
a=a_0e^{\{\mu_0t+(\mu_5+0.5\mu_3)\cosh^{-2}[t]-0.25\mu_5\cosh^{-4}[t]\}}\cosh^{(\mu_1+\mu_3+\mu_5)}[t].
\end{align}
Note that if $\mu_5=-\mu_1-\mu_3$ then the scale factor becomes
\begin{align}
a=a_0e^{\{\mu_0t-(\mu_1+0.5\mu_3)\cosh^{-2}[t]+0.25(\mu_1+\mu_3)\cosh^{-4}[t]\}}.
\end{align}
 In general as $t \rightarrow \infty$ 
\begin{align}
a\rightarrow a_0e^{(\mu_0+\mu_1+\mu_3+\mu_5)t}.
\end{align}
We need in the following formulas
\begin{align}
	\dot{H}=\mu_1+(3\mu_3-\mu_1)y^2+(5\mu_5-3\mu_3)y^4-5\mu_5y^6,
\end{align}
\begin{align}
	H^2=\mu_0^2+2\mu_0\mu_1y+\mu^2_1y^2+2\mu_0\mu_3y^3+2\mu_1\mu_3y^4+2\mu_0\mu_5y^5+(\mu_3^2+2\mu_1\mu_5)y^6+2\mu_3\mu_5y^8+\mu_5^2y^{10}.
\end{align}
Then Eq.(3.4) gives
\begin{align}
	\nu_0: &=&-(2\mu_1+3\mu_0^2),\\
	\nu_1: &=&-6\mu_0\mu_1,\\
	\nu_2: &=&-2(3\mu_3-\mu_1)-3\mu_1^2,\\
	\nu_3: &=&-6\mu_0\mu_3,\\
	\nu_4: &=&6\mu_3(1-\mu_1)-10\mu_5,\\
	\nu_5: &=&-6\mu_0\mu_5,\\
	\nu_6: &=&10\mu_5-3\mu_3^2-6\mu_1\mu_5,\\
	\nu_7: &=&0,\\
	\nu_8: &=&-6\mu_3\mu_5,\\
	\nu_9: &=&0,\\
	\nu_{10}: &=&-3\mu_5^2.
	\end{align}
After some algebra  we get the following important expressions
\begin{align}
2XK_X=-2[\mu_1+(3\mu_3-\mu_1)y^2+(5\mu_5-3\mu_3)y^4-5\mu_5y^6],
\end{align}
$$
X=\frac{a_0^6}{\kappa}[\mu_1+(3\mu_3-\mu_1)y^2+(5\mu_5-3\mu_3)y^4-
$$
\begin{align}
5\mu_5y^6]^2\cosh^{6[\mu_1+\mu_3+\mu_5)}[t]e^{\{6[\mu_0t+(\mu_5+0.5\mu_3)
\cosh^{-2}[t]-0.25\mu_5\cosh^{-4}[t]]\}}.
\end{align}
Hence we can recover the function $\phi$:
\begin{align}
\phi=\partial^{-1}_t\sqrt{-2X}.
\end{align}

Finally we present the formulas  for the density of energy and pressure:
\begin{align}
\rho_k=3(\mu_0+\mu_1y+\mu_3y^3+\mu_5y^5)^2,
\end{align}
\begin{align}
p_k=\nu_0+\nu_1y+\nu_2y^2+\nu_3y^3+\nu_4y^4+\nu_5y^5+\nu_6y^6+\nu_8y^8	+\nu_{10}y^{10}.
\end{align}

\section{Cosmic speed-up}
Now we would like to show that the above presented models at least for some values of the parameters $\mu_j$ admit the cosmic acceleration. We illustrate our arguments for Example 2 ($M=6$).  In this case we have 
\begin{align}
a|_{t \rightarrow \infty}\rightarrow a_0e^{(\mu_0+\mu_1+\mu_3)t}.
\end{align}
The corresponding acceleration is
\begin{align}
\ddot{a}|_{t \rightarrow \infty}\rightarrow (\mu_0+\mu_1+\mu_3)^2a_0e^{(\mu_0+\mu_1+\mu_3)t}.
\end{align}
Hence follows that  $\ddot{a}\succ0$ so we have the acceleration phase. The exception is  the case when $\mu_0+\mu_1+\mu_3=0$ that corresponds to the transition between the deceleration and acceleration phases.

For the equation of state we get
\begin{align}
w_k=\frac{p_k}{\rho_k}=-1+\frac{2XK_X}{2XK_X-K}=-1-\frac{2\mu_1-(2\mu_1-6\mu_3)\tanh^2[t]-6\mu_3\tanh^4[t]}{3(\mu_0+\mu_1\tanh[t]+\mu_3\tanh^3[t])^2}.
\end{align}
Hence we see that $w_k|_{t\rightarrow 0}=-1-\frac{2\mu_1}{3\mu_0^2}, \quad w_k|_{t\rightarrow\infty}=-1$. Here  can be observed the last two phases 
\begin{figure}[h]
	\centering
		\includegraphics{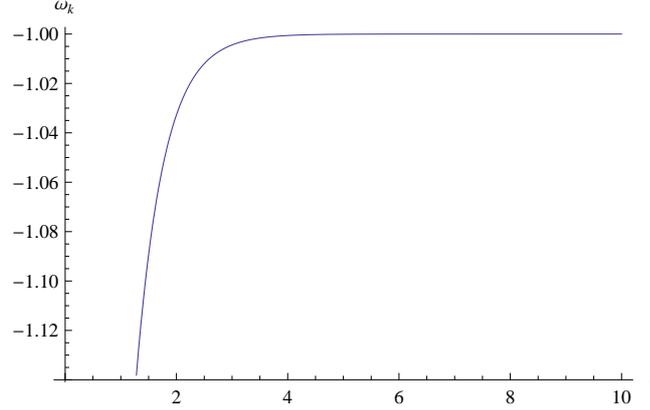}
		\caption{The curve corresponds to the equation of state $w_k$ as a function of $t$ with $\mu_j=1, a_0=-\kappa=1$.}
	\label{fig:ot}
\end{figure}\\
of the evolution of the universe: radiation dominated ($w=\frac{1}{3}$) if $\mu_1=-2\mu_0^2$ or matter dominated ($w=0$) if $\mu_1=-\frac{3}{2}\mu_0^2$ and late acceleration ($w=-1)$.  Figure 1 shows the evolution of the equation of state $w_k$ as a function of $t$ with $\mu_j=1, a_0=-\kappa=1$. We mention that in our natural units, constants $\nu_j$ and $\mu_j$ are dimensionless. 

\begin{figure}[h]
	\centering
		\includegraphics{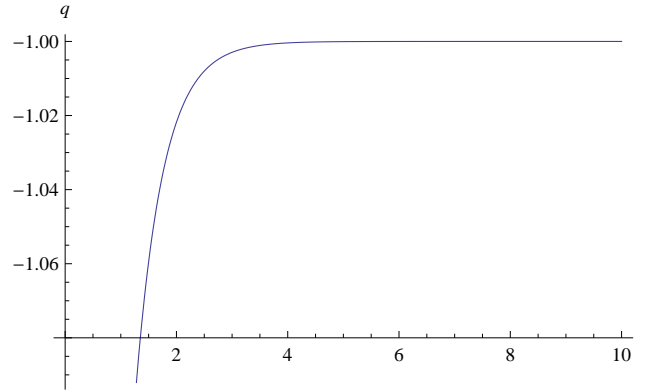}
			\caption{The curve corresponds to the deceleration parameter $q$ as a function of $t$ with $\mu_j=1, a_0=-\kappa=1$.}
	\label{fig:qt}
\end{figure}
At the same time, the deceleration parameter takes the form 

\begin{align}
q=-\frac{\ddot{a}a}{\dot{a}^2}=-1-\frac{\dot{H}}{H^2}=-1-\frac{\mu_1+(3\mu_3-\mu_1)\tanh^2[t]-3\mu_3\tanh^4[t]}{(\mu_0+
\mu_1\tanh[t]+\mu_3\tanh^3[t])^2}.\end{align}
Hence we see that $q|_{t\rightarrow 0}=-1-\frac{\mu_1}{\mu_0^2}, \quad q|_{t\rightarrow\infty}=-1$. So we find that our model predicts that the transition from deceleration to acceleration occurs at $\mu_1=-\mu_0^2$ in early time.

Figure 2 shows the evolution of the deceleration parameter  $q$ as a function of $t$ for our model (Example 2) with the $\mu_j=1, a_0=-\kappa=1$.
Also we note that $\rho_k|_{t\rightarrow 0}=3\mu_0^2, \quad p_k|_{t\rightarrow 0}=-(2\mu_1+3\mu_0^2)$ and $\rho_k|_{t\rightarrow \infty}=3(\mu_0+\mu_1+\mu_3)^2, \quad p_k|_{t\rightarrow\infty}=-3(\mu_0+\mu_1+\mu_3)^2$.

\section{Conclusion}
In the present paper,  we have investigated  the following model  of k-essence: \begin{align}
	K=\nu_0(t)+\nu_2(t)\tanh[t]^2+\nu_3(t)\tanh[t]^3+ ... +\nu_M(t)\tanh[t]^M
\end{align} for the cases when $\nu_j=const$ and $M=2, 6, 10$.  These k-essence models were presented in the parametric forms. The  exact analytical solutions of the corresponding equations of motion are found. We studied their cosmological behaviors. It is shown that they can give rise to cosmic acceleration. 
\section{Acknowledgement}
We would like to thank the anonymous referee for useful remarks and suggestions that allowed us to improve the paper.


\begin{thebibliography}{99}  
\bibitem{Ries} A.G. Riess et al. [Supernova Search Team Collaboration],  Astron. J. \textbf{116}, 1009 (1998); S.Perlmutter et al. [Supernova Cosmology Project Collaboration],  Astrophys. J. \textbf{517}, 565 (1999) 
\bibitem{1} S. Weinberg,  Rev. Mod. Phys. \textbf{61}(1989) 1;
V. Sahni and A. A. Starobinsky,  Int. J. Mod. Phys. D \textbf{9}(2000) 373;
S. M. Carroll,  Living Rev. Rel. \textbf{4}, 1 (2001) ;
E. J. Copeland, M. Sami and S. Tsujikawa,  Int. J. Mod. Phys. D \textbf{15}, (2006) 1753
\bibitem{2} R.-J. Yang and S. N. Zhang, Mon. Not. R. Astron. Soc. \textbf{407} (2010) 1835-1841.
\bibitem{3} C. Armendariz-Picon, V. Mukhanov, and Paul J. Steinhardt,  Phys. Rev. Lett. \textbf{85} (2000) 4438 
\bibitem{4} C. Armendariz-Picon, T. Damour, and V. Mukhanov,  Phys. Lett. B \textbf{458} (1999) 209 
\bibitem{5} M. Malquarti, E. J. Copeland, A. R. Liddle, and M. Trodden,  Phys. Rev. D \textbf{67} (2003) 123503
\bibitem{6} J. M. Aguirregabiria, L. P. Chimento, and R. Lazkoz,  Phys. Rev. D \textbf{70} (2004) 023509
\bibitem{7} L. R. Abramo and N. Pinto-Neto,  Phys. Rev. D \textbf{73} (2006) 063522
\bibitem{8} A. D. Rendall,  Class. Quantum Grav. \textbf{23} (2006) 1557 
\bibitem{9} R. Das, T. W. Kephart, and R. J. Scherrer, 
Phys. Rev. D \textbf{74} (2006) 103515 
\bibitem{10} T. Chiba, S. Dutta, and Robert J. Scherrer,  Phys. Rev. D \textbf{80} (2009) 043517 
\bibitem{11} N. Cruz, P. F. Gonzalez-Diaz, A. Rozas-Fernandez, and G. Sanchez,  Phys. Lett. B
\textbf{679} (2009) 293 
\bibitem{12} L. N. Granda and A. Oliveros, \textit{Holographic reconstruction of the k-essence and dilaton models}, [arXiv:0901.0561]
\bibitem{13} X.-T. Gao and R.-J. Yang,  Phys. Lett. B \textbf{687} (2010) 99 
\bibitem{14} F. Arroja and M. Sasaki,  Phys. Rev. D \textbf{81} (2010) 107301 
\bibitem{15} A. Diez-Tejedor and A.  Int. J. Mod. Phys. D \textbf{14} (2005) 1561 
\bibitem{16} L. P. Chimento, M. Forte, and R. Lazkoz,  Mod. Phys.
Lett. A. \textbf{20} (2005) 2075 
\bibitem{17} S. Sur and S. Das, \textbf{JCAP 01} (2009) 007 
\bibitem{18} L. P. Chimento and R. Lazkoz,  Phys. Lett. B \textbf{639},
(2006) 591 
\bibitem{19} N. Bilic,  Phys. Rev. D \textbf{78} (2008) 105012 
\bibitem{20} L. P. Chimento,  Phys. Rev. D \textbf{69} (2004) 123517
\bibitem{21} R. J. Scherrer, Phys. Rev. Lett. \textbf{93} (2004) 011301 
\bibitem{22} D. Bertacca, S. Matarrese, and M. Pietroni,  Mod. Phys. Lett. A \textbf{22}
(2007) 2893 
\bibitem{23} N. Bose and A. S. Majumdar,  Phys. Rev. D \textbf{79} (2009)
103517 
\bibitem{24} A. Diez-Tejedor and A. Feinstein,  Phys. Rev. D
\textbf{74} (2006) 023530 
\bibitem{25} L. P. Chimento and A. Feinstein,  Mod. Phys. Lett. A \textbf{19} (2004) 761
\bibitem{26} R.-J. Yang and X.-T. Gao, Chin. Phys. Lett.
\textbf{26} (2009) 089501.
\bibitem{27} R. J. Yang, S. N. Zhang, and Y. Liu,  J.
Cosmol. Astropart. Phys. \textbf{01} (2008) 017
\bibitem{28} R.-J. Yang and S. N. Zhang, Chin. Phys. Lett. \textbf{25} (2008) 344.
\bibitem{29} A. A. Sen,  \textbf{JCAP 03} (2006) 010 
\bibitem{30} R. de Putter and E. V. Linder,  Astropart. Phys. \textbf{28} (2007) 263 
\bibitem{31} R.-J. Yang and X.-. Gao, \textit{Restrictions on purely kinetic k-essence}, [arXiv:1005.4343].
\bibitem{32} J. Garriga and V. F. Mukhanov,  Phys. Lett. B \textbf{458} (1999) 219 
\bibitem{33} M. Gasperini and G. Veneziano,  Phys. Rept. \textbf{373} (2003) 1
\bibitem{34} R. Bean, \textit{TASI Lectures on Cosmic Acceleration}, [arXiv:1003.4468].
\bibitem{35} S. Tsujikawa, \textit{Dark energy: investigation and modeling}, [arXiv:1004.1493].
\bibitem{36} S. Unnikrishnan and L. Sriramkumar, , Phys. Rev. D \textbf{81} (2010) 103511 
\bibitem{37} A. J. Christopherson and K. A. Malik,  Phys. Lett. B \textbf{675} (2009)
159 
\bibitem{38} E. Babichev, V. Mukhanov and A. Vikman, 
JHEP \textbf{02} (2008) 101 
\bibitem{39} Jean-Philippe Bruneton,  Phys. Rev. D \textbf{75} (2007) 085013 
\bibitem{40} J. U. Kang, V. Vanchurin, and S. Winitzki,  Phys.
Rev. D \textbf{76 }(2007) 083511 
\bibitem{41} C. Bonvin, C. Caprini, and R. Durrer,  Phys. Rev. Lett. \textbf{97 }(2006) 081303
\bibitem{42} V. Gorini, A. Y. Kamenshchik, U. Moschella, O. F. Piattella, and A. A. Starobinsky,  JCAP \textbf{02} (2008) 016 
\bibitem{43} G. Ellis, R. Maartens, and M. MacCallum,  Gen. Rel. Grav. \textbf{39} (2007) 1651
\bibitem{44} E.Silverstein, D.Tong, Phys. Rev. D \textbf{70} (2004) 103505; M.Alishahiha, E.Silverstein, D.Tong, Phys. Rev. D \textbf{70} (2004) 123505 
\bibitem{45} S. Nojiri,  \textit{Towards the unification of late-time acceleration and inflation by k-essence model}, [arXiv:1006.5201]; 
\bibitem{46} E. J. Copeland, A. R Liddle, and D. Wands,  Phys. Rev. D \textbf{57}
(1998) 4686 
\bibitem{47} P. G. Ferreira and M. Joyce, Phys. Rev. Lett. \textbf{79} (1997) 4740
\bibitem{48} E. J. Copeland, M. Sami, and S. Tsujikawa,  Int. J. Mod. Phys.D \textbf{15} (2006) 1753

\bibitem{49} X.-m. Chen, Y. Gong, and E. N. Saridakis,  JCAP \textbf{04} (2009) 001
\bibitem{50} Kei-ichi Maeda and Y. Fujii,  Phys. Rev. D \textbf{79} (2009)
084026 
\bibitem{51} J. M. Aguirregabiria and Ruth Lazkoz, Phys. Rev. D \textbf{69} (2004) 123502
\bibitem{Yang1} R.-J. Yang, X.-T. Gao,  \textit{Phase-space analysis of k-essence cosmology}. [arXiv:1006.4986]
\bibitem{MN} J. Matsumoto, S. Nojiri,    Phys. Lett. B \textbf{687} (2010) 236  
\bibitem{BBM} D. Bertacca, N. Bartolo, S. Matarrese,  \textit{Unified Dark Matter Scalar Field Models},  [arXiv:1008.0614]
\bibitem{N1} S. Nojiri, S.D.  Odintsov, AIP Conf. Proc. 
 \textbf{1115} (2009) 212 
\bibitem{Eli1} E. Elizalde,  S. Nojiri,  S.D. Odintsov,  D. Saez-Gomez, V. Faraoni, 
Phys. Rev. D \textbf{77}, 106005 (2008).
\bibitem {MR1}  R. Myrzakulov,  D. Saez-Gomez, A. Tureanu, \textit{On the $\Lambda CDM$ Universe in $f(G)$ gravity}, [arXiv:1009.0902]
\bibitem {MR2} E. Elizalde, R. Myrzakulov, V.V. Obukhov, D. Saez-Gomez, Classical and Quantum Gravity \textbf{27},  095007 (2010) 
\end{thebibliography}
\end{document}